\documentclass[sigconf]{acmart}
\usepackage{enumitem}
\setlist{listparindent=1.25em}
\usepackage[figuresright]{rotating}
\usepackage{tabularx}
\usepackage{balance}
\usepackage{tikz}
\usepackage{framed}
\usepackage{multicol}
\usepackage{pgfplots,pgfplotstable}
\pgfplotsset{compat=1.18}
\usepackage{tikzviolinplots}
\usetikzlibrary{shapes.geometric, arrows, positioning, patterns, pgfplots.statistics}

\newcommand\identifying[1]{#1}

\AtBeginDocument{%
  \providecommand\BibTeX{{%
    \normalfont B\kern-0.5em{\scshape i\kern-0.25em b}\kern-0.8em\TeX}}}

\acmConference[arXiv Pre-print]{arXiv Pre-print}{}{}
\acmBooktitle{arXiv Pre-print}
\renewcommand\footnotetextcopyrightpermission[1]{}
\makeatletter
\renewcommand\@formatdoi[1]{\ignorespaces}
\makeatother

\begin{document}
\fancyfoot{}
\fancyhead{}
\title{Writing Blog Posts Helps Students Connect Experiential Learning to the Workplace}
\pagestyle{plain}
\author{Utsab Saha}
\email{usaha@csumb.edu}
\orcid{0009-0004-0458-8697}
\affiliation{
 \institution{Computing Talent Initiative}
 \city{Marina}
 \state{California}
 \country{USA}
}

\author{Lola Egherman}
\email{lola@codeday.org}
\orcid{0009-0005-0684-7995}
\affiliation{
 \institution{CodeDay}
 \city{St. Paul}
 \state{Minnesota}
 \country{USA}
}

\author{Ramiz Rahman}
\email{ramiz@codeday.org}
\orcid{0009-0007-7876-0172}
\affiliation{
 \institution{CodeDay}
 \city{West Bengal}
 \country{India}
}

\author{Mohd Toukir Khan}
\email{toukir@codeday.org}
\orcid{0009-0003-2394-1655}
\affiliation{
 \institution{CodeDay}
 \city{Chhattisgarh}
 \country{India}
}

\author{Kevin Wang}
\email{kevin@mentorsintech.com}
\orcid{0009-0003-4227-7745}
\affiliation{
 \institution{Mentors in Tech}
 \city{Bellevue}
 \state{Washington}
 \country{USA}
}

\author{Tyler Menezes}
\email{tylermenezes@codeday.org}
\orcid{0000-0002-7975-2533}
\affiliation{
 \institution{CodeDay}
 \city{Seattle}
 \state{Washington}
 \country{USA}
}

\renewcommand{\shortauthors}{Saha et al.}

\begin{abstract}
Undergraduates in work-based learning experiences often produce meaningful contributions as viewed by their supervisors, yet r\-eport a negative perception of their contributions because they struggled during the process or produced only a few lines of code change. As a result, many omit these contributions from their resumes and job interviews, losing a meaningful signal of technical ability.

This study examines how guided blog posts help CS students in work based learning experiences reflect on what they learned and contextualize their experiences. It also evaluates the depth of reflection produced.

The study included twenty-five juniors and seniors studying CS at CTCs and other affordable local colleges. All participated in one cohort during Fall 2024.

Each student was assigned a simple open source issue to solve from a popular open source project over the course of several weeks with the help of an industry mentor. While working on the project, students drafted a LinkedIn blog post using a five-section outline covering project mission, assigned issue, technical architecture, ch\-allenges faced, and submitted solution. We conducted a thematic analysis of the published posts and measured reflection depth using Mejia and Turns's Knowledge Gain instrument.

Four themes emerged from the posts: identifying problem solving techniques, growth mindset, the challenges and benefits of collaborative development, and the impacts of their contribution on users. Additionally, students demonstrated deep reflection across all four Knowledge Gain constructs.

Structured blog posts offer a low-cost addition to experiential CS learning such as capstones, micro-internships, internships, and apprenticeships. This study is descriptive; future work should compare outcomes against a control group.
\end{abstract}

\maketitle
\section{Introduction}

Experiential learning programs in computer science education face a persistent challenge: while students often complete impressive technical work, they frequently struggle to recognize its value and articulate their experiences to potential employers. This disconnect between achievement and self-perception can diminish the career-advancement benefits that such programs aim to provide. \cite{ionsBarriersConstructingExperiential2019,duroseLostTranslationPreparing2016,jacksonEncouragingStudentsDraw2019}

The authors have operated \identifying{CodeDay Labs}, a work-based learning program where undergraduate computer science students contribute to popular open source software projects. Over multiple cohorts, students have successfully submitted thousands of pull requests to major projects—accomplishments that consistently impress professional software engineers. \identifying{\cite{menezesOpenSourceInternshipsIndustry2022,narayananScalableApproachSupport2023,hangIndustryMentoringInternship2024}} However, post-program feedback revealed that many students felt negatively ab\-out their experience because they encountered difficulties during the process or produced what they perceived as minimal code cha\-nges. Consequently, students often omitted these significant achie\-vements from their resumes and failed to discuss them during job interviews, missing valuable opportunities to demonstrate real software development experience.

To address this gap between accomplishment and self-percept\-ion, we sought to implement a structured reflection exercise requiring students to write blog posts about their open source contributions on LinkedIn. This approach was designed to help students re-contextualize their technical work within a professional narrative while simultaneously engaging their emerging professional networks. Students followed a provided outline covering the project's mission and impact, technical architecture, their assigned issue, problem-solving process, and final results. They developed these posts incrementally throughout the program, publishing upon completion of their contributions.

This paper presents a mixed-methods evaluation of this intervention. Our research questions were: \textbf{(RQ1)} within the provided framework, what topics do students reflect on; and \textbf{(RQ2)} how thoroughly do the structured blog posts allow students to reflect on their experience.

To answer these questions, we conducted a thematic analysis of student blog posts to identify patterns in reflection and learning articulation, complemented by a quantitative survey measuring the depth of student reflection.

Our findings demonstrate that structured blog writing improved students' ability to recognize and communicate the value of their experiential learning. This approach has implications for enhancing other work-based learning programs, including capstones, internships, micro-internships, and apprenticeships, by providing a scalable method for fostering meaningful reflection and professional communication skills.
\section{Background}

\subsection{The Challenge of Experiential Learning}

Experiential learning has become increasingly recognized as essential in computer science education, providing students with authentic problem-solving experiences that bridge the gap between theoretical knowledge and practical application. Research by the Burning Glass Institute and Strada Education Foundation indicated that 30\% of computer science graduates in the United States are underemployed, but that students who had an internship were much more likely to have college-level employment. \cite{stradaTalentDisrupted2024} Through internships, capstone projects, and open source contributions, students develop technical skills while navigating the complexities of real-world software development environments.

However, a persistent challenge is the disconnect between students' technical accomplishments and their ability to recognize and articulate the professional value of their work. \cite{duroseLostTranslationPreparing2016} Students often focus narrowly on the immediate technical outcome -- such as the number of lines of code written or the complexity of algorithms implemented -- while overlooking the broader professional competencies they develop through the problem-solving process, collaboration with distributed teams, and navigation of unfamiliar codebases. \cite{cropleyPromotingCreativityInnovation2015,towersStudentsPerceptionsEngineering2011,egglestonEmployerStudentMismatch2022} This leads to undervaluing significant learning experiences, particularly when students encounter typical software development challenges like debugging complex issues, understanding legacy code, or working within established development workflows. The result is that students may dismiss meaningful professional experiences as failures or insignificant contributions, failing to leverage these experiences effectively in job applications, interviews, and career development conversations with potential employers. \cite{jacksonEncouragingStudentsDraw2019,duroseLostTranslationPreparing2016}

Students face several challenges in recognizing the value of experiential learning and articulating their experiences to employers. These include difficulties in constructing reflective narratives, articulating tacit knowledge, and accurately recalling past learning experiences. \cite{ionsBarriersConstructingExperiential2019} In addition to simply providing opportunities for practical application of theory, researchers have suggested to develop students' reflective skills to address these issues. \cite{gorkaDevelopingRealisticCapstone2007,frenchMountaintopCorporateLadder2015}

\subsection{Professional Identity Development Through Reflective Writing}

When properly implemented, reflection serves as a key ingredient in transforming students' experiences into meaningful learning \cite{marienauBringingStudentsExperience2002} and has long been considered essential for continual professional growth in fields like medicine and education. \cite{robertsonReflectionProfessionalPractice2005}

Reflection plays a crucial role in transforming experience into professional growth for educators. It helps individuals evaluate experiences, learn from mistakes, and revise practices; facilitates the connection between theory and practice; and can encourage deep learning and behavior change. \cite{lockyerKnowledgeTranslationRole2004,swainStudyingTeachersTransformations1998} However, reflection requires time, effort, and personal investment. \cite{chalikandyReflectionToolProfessional2014} 

Structured writing can be a powerful tool for reflection, enhancing critical thinking and personal growth. Various approaches have been developed to foster reflective writing, including structured reflective logs, guided journal writing, and frameworks with specific reflective acts. \cite{sarigFosteringReflectiveWriting2005, kolarPreserviceGeneralEducators2002} These structured methods help learners move from descriptive to analytical reflection. \cite{samuelsCrossingThresholdDescription2007}

\subsection{Reflective Writing in Computing Education}

Reflection in computing education has received relatively little interest compared to reflection in education research as a whole.

Computing education researchers have found that reflective writing can help strengthen metacognition and self‑regulated learning. For example, graduate computational science courses that embed journals and reflective prompts (e.g., "how did I approach this problem?") report gains in metacognitive awareness, self‑regulated learning, problem‑solving, and critical thinking. \cite{zarestkyReflectiveWritingSupports2022,sudirmanReinforcingReflectivePractice2024}

Reflection may also support computational thinking and conceptual understanding. In one study of reflection in CS‑unplugged activities, two‑column notes and daily diaries led to better computational thinking skills than unplugged activities alone. \cite{ugurImplementingReflectiveThinking2024} Another study provided evidence that short reflection prompts just before exams in a Computer Organization course may have improved exam performance and helped students develop a holistic, "big picture" understanding. \cite{reschAnalysisEffectAnswering2022}

Computing educators have also observed many of the same benefits of reflection to professional identity development that were described in the general education literature. \cite{langerReflectingPracticeUsing2002,dicksonBringingReflectionComputer2019,georgeLearningReflectiveJournal2002}

\subsection{Theoretical Framework}

We evaluated a number of theoretical frameworks to use in structuring this work, \cite{rogersReflectionHigherEducation2001,kreberAnalysisTwoModels2004} and felt the framework proposed by Turns et al. in "Integrating Reflection into Engineering Education" was best suited to our research goals. \cite{turnsIntegratingReflectionEngineering2014} They define reflection as an intentional and dialectical thinking process where individuals revisit specific features of their experiences—those aspects they are consciously aware of and can describe—and apply various lenses (such as subject matter knowledge, identity understanding, or social perspectives) to interpret these features and derive meaningful understanding. This meaning-making process creates tension between using existing interpretive frameworks and revising them based on new experiential insights, ultimately guiding future actions and experiences. The framework emphasizes that effective reflection requires deliberate engagement with this process, moving beyond implicit meaning-making to conscious analysis that connects past experience to future behavior and learning.
\section{Model}
\begin{figure*}
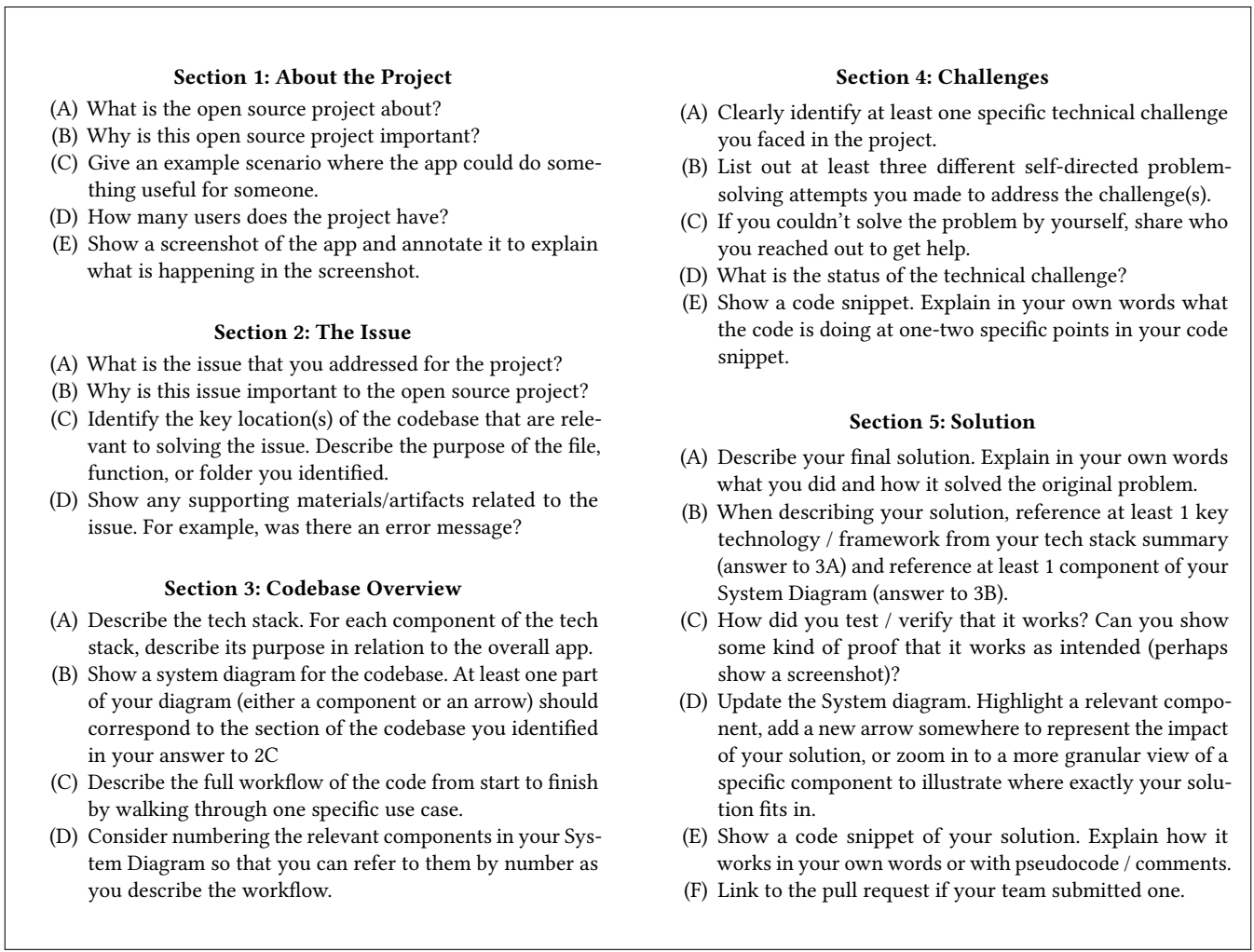

    \renewcommand{\theenumi}{\Alph{enumi}}
    \begin{framed}
        \begin{multicols}{2}
            \textbf{Section 1: About the Project}
            \begin{enumerate}
                \item What is the open source project about?
                \item Why is this open source project important?
                \item Give an example scenario where the app could do something useful for someone.
                \item How many users does the project have?
                \item Show a screenshot of the app and annotate it to explain what is happening in the screenshot.
            \end{enumerate}
            \vspace{12pt}
            \textbf{Section 2: The Issue}
            \begin{enumerate}
                \item What is the issue that you addressed for the project?
                \item Why is this issue important to the open source project?
                \item Identify the key location(s) of the codebase that are relevant to solving the issue. Describe the purpose of the file, function, or folder you identified.
                \item Show any supporting materials/artifacts related to the issue. For example, was there an error message?
            \end{enumerate}
            \vspace{12pt}
            \textbf{Section 3: Codebase Overview}
            \begin{enumerate}
                \item Describe the tech stack.  For each component of the tech stack, describe its purpose in relation to the overall app.
                \item Show a system diagram for the codebase.  At least one part of your diagram (either a component or an arrow) should correspond to the section of the codebase you identified in your answer to 2C
                \item Describe the full workflow of the code from start to finish by walking through one specific use case.
                \item Consider numbering the relevant components in your System Diagram so that you can refer to them by number as you describe the workflow.
            \end{enumerate}
            \vspace{12pt}
            \textbf{Section 4: Challenges}
            \begin{enumerate}
                \item Clearly identify at least one specific technical challenge you faced in the project.
                \item List out at least three different self-directed problem-solving attempts you made to address the challenge(s).
                \item If you couldn't solve the problem by yourself, share who you reached out to get help.
                \item What is the status of the technical challenge?
                \item Show a code snippet.  Explain in your own words what the code is doing at one-two specific points in your code snippet.
            \end{enumerate}
            \vspace{12pt}
            \textbf{Section 5: Solution}
            \begin{enumerate}
                \item Describe your final solution.  Explain in your own words what you did and how it solved the original problem.
                \item When describing your solution, reference at least 1 key technology / framework from your tech stack summary (answer to 3A) and reference at least 1 component of your System Diagram (answer to 3B).
                \item How did you test / verify that it works?  Can you show some kind of proof that it works as intended (perhaps show a screenshot)?
                \item Update the System diagram.  Highlight a relevant component, add a new arrow somewhere to represent the impact of your solution, or zoom in to a more granular view of a specific component  to illustrate where exactly your solution fits in. 
                \item Show a code snippet of your solution.  Explain how it works in your own words or with pseudocode / comments.
                \item Link to the pull request if your team submitted one.
            \end{enumerate}
        \end{multicols}
    \end{framed}
    \caption{Blog post template that was provided to students}
    \label{fig:blogPostTemplate}
\end{figure*}

\subsection{Empirical Setting}

This work was situated in \identifying{CodeDay Labs}, an internship-style program where students contribute to open source software projects. \identifying{\cite{menezesOpenSourceInternshipsIndustry2022}} In the program, students work individually or in small teams of 2-4 members to address issues in established open source repositories, applying their computer science education in authentic software development contexts. All contributions are made to projects with Open Source Initiative approved licenses,\cite{OpenSourceDefinition} ensuring student work becomes part of the broader open source ecosystem. The program can operate as part of a capstone course \identifying{\cite{hangIndustryMentoringInternship2024}} or as an extracurricular. \identifying{\cite{narayananScalableApproachSupport2023}}

Each cohort runs for 4 or 12 weeks, beginning with a one-week onboarding process that introduces core software engineering practices. Students in the 4-week program are matched with extremely simple initial issues—such as fixing failing tests, improving error messages, or adding translations—to provide an accessible entry point into open source contribution, while students in the 12-week program are matched with more complex issues such as adding a new feature. Students participate in weekly team meetings with an industry mentor and structured agendas to guide their progress toward completing a pull request. Students also individually answer twice-weekly asynchronous "stand-ups" asking them to reflect on what they accomplished and plan to work on next. \identifying{\cite{menezesAIGradingStandupUpdates2024}}

The program leverages industry-standard collaboration tools su\-ch as GitHub for code review and issue tracking, Slack for team communication, and video conferencing for synchronous meetings. \cite{jacksonCollaborationToolsDevelopers2022} Mentoring also plays a crucial role in helping students become successful open source contributors. \cite{fengMultifacetedNatureMentoring2025} Accordingly, students receive mentorship focused on developing the confidence to work on complex problems independently from professional software engineers who are \textit{not} subject matter experts, as well as highly specific technical mentorship from project maintainers. Teaching Assistants provide additional mentorship and support in debugging, allowing the program to scale mentor participation while maintaining quality guidance.

\subsection{Blog Post Writing Process}

Using the backward design process described by Wiggins and Mc\-Tighe, \cite{wigginsUnderstandingDesign2008} we created a suggested blog post template designed to guide students' reflection. (Figure \ref{fig:blogPostTemplate}) Students were provided with the template at the beginning of the program, and were instructed to write a section of the blog post each week. The sections were structured so that they likely corresponded to what students were working on at that time. For example, week 1 of the experience was designed for students to learn more about their project, and Section 1 of the blog post focuses on information about the project. By week 4, students have typically encountered challenges, and the blog post asks them to reflect on those challenges. Students were required to submit their blog post updates each week, although the content was not graded and feedback was not provided.

Some students participated in a longer program, in which they contributed to open source for up to 12 weeks. These students were still encouraged to work on the blog post each week, but were encouraged to go back to revise the blog post each week.

Students were strongly encouraged to publish their blog posts on LinkedIn. We chose LinkedIn because research suggests it provides an affordable platform for students to showcase their skills and connect with potential employers, and because it is the fastest-growing platform among college students and recent graduates for professional networking and job searches. \cite{carmackUsingTheoryPlanned2018} Previous work has found that many students struggle to effectively market themselves on the platform and may not fully understand its potential, \cite{danielsLinkedInBlundersMixed2023} and we hoped that the blog posts might help students attract additional attention from professional connections. Due to length constraints, however, the evaluation of potential impact on students' job prospec\-ts is beyond the scope of this paper.
\section{Methods}

This study employed a mixed-methods approach to evaluate the effectiveness of blog post writing as a reflective tool for students participating in open source software contributions. Data collection consisted of two primary components: thematic analysis of student blog posts and a quantitative survey measuring reflection depth.

Participants in this study were drawn from a single 4-week cohort of the \identifying{CodeDay Labs} program during Fall 2024, all of whom were juniors and seniors pursuing 4-year BS degrees in computer science at state-funded universities and community and technical colleges in California and Washington. All students in the cohort were required to work on the blog post as part of the program.

\subsection{Thematic Analysis}

We conducted a qualitative thematic analysis of student blog posts (n=25) using the ATLAS.ti software. Blog posts largely followed the structured outline shown in Figure \ref{fig:blogPostTemplate}; this consistent structure facilitated systematic analysis across posts.

The coding process followed Braun and Clarke's six-phase framework for thematic analysis. \cite{braunUsingThematicAnalysis2006} Although Braun and Clarke's original framework encompasses a range of approaches, the method we employed most closely aligns with a coding reliability approach, in which independent coders apply a shared codebook and measure agreement. This contrasts with reflexive thematic analysis, which treats coding as an iterative, researcher-driven process without formal inter-rater reliability measures. \cite{braunOneSizeFits2021} We chose the coding reliability approach because the structured nature of the blog post template made consistent code application across coders feasible, and because measuring agreement strengthened our confidence in the identified themes.

 Two independent coders initially developed a preliminary codebook through open coding of a subset of blog posts (n=10). The codebook was iteratively refined through discussion and consensus-building until saturation was reached. Codes captured both explicit reflective statements and implicit learning demonstrations within student narratives.

To ensure coding reliability, we calculated inter-rater reliability at the theme level using Cohen's Kappa coefficient using 20\% of the dataset. The overall Cohen's Kappa was $\kappa = 0.78$, indicating substantial agreement between coders. \cite{landisMeasurementObserverAgreement1977} Discrepancies were resolved through discussion, and the remaining posts were divided between coders for analysis.

Themes were identified through pattern recognition across codes, focusing on evidence of reflection depth, learning outcomes, and professional identity development. We specifically examined how students articulated connections between their technical struggles and professional software development practices.

\subsection{Knowledge Gain Through Reflection Survey}

To quantitatively assess the value of blog post writing as a reflection tool, we administered the Knowledge-Gain Survey instrument developed by Mejia and Turns. \cite{mejiaCreatingCapacityExplore2021} The instrument consists of 16 items across four constructs: (1) Engineering Self, (2) Course Understandings, (3) Areas for Growth, and (4) Social Impact. Items use a 5-point Likert scale ranging from "strongly disagree" to "strongly agree." (Maximum scores for each construct ranged from 12-20; for ease of interpretation, we normalized the scores to 1-5.)

The survey was distributed to all students who participated in the program and published a blog post (regardless of whether they completed an open source contribution or if it was accepted by the project).
\section{Findings}

\subsection{(RQ1) Within the provided framework, what topics do students reflect on?}

\subsubsection{Identifying problem solving techniques}

The most prevalent theme, appearing in 24 of 25 posts, was the identification of problem-solving techniques. Students described techniques they used to solve complex problems when they didn't immediately know the answer: reading documentation, looking for examples in related files, narrowing down the problem through debugging and experimentation, and asking for help.

\vspace{12pt}

\textit{"The last thing I needed to figure out was what to specifically place in the two files. To figure this out, I did a look through the project maintainers' other repositories for other examples of how they made their composer files."}

\vspace{12pt}

\textit{"One attempt I made to address this challenge was carefully reading through the build instructions for developers."}

\vspace{12pt}

\textit{"When I contacted my teammates, I found that [teammate] had faced the same problem. However, it turned out this was a codespaces issue, not a Docker issue."}

\vspace{12pt}

\textit{"Relevant to the issue of generating an RSA key, I had to identify key locations to where the codebase is generating an RSA key by pressing a button."}

\vspace{12pt}

\textit{"The error messages in the terminal provided some guidance on where the issues lay. I attempted to resolve each error by tweaking the database connection string format. I tried changing various values in the .env.local file and re-running prisma db pull to see if the connection errors persisted."}

\subsubsection{Reward for perseverance/growth mindset}

In 19 of 25 posts, students spoke proudly of how persistent, determined efforts at problem solving were rewarded by the answer to a problem "clicking" or the bigger picture coming into view.

\vspace{12pt}

\textit{"A third attempt I made was actually building the project and using Docker in the process. Doing this really helped me understand why Docker was being used in this project. I now understand what Docker is and how it is used."}

\vspace{12pt}

\textit{"While I was waiting for a response, I began reading through the JavaScript files and I noticed there was a distinction between the test cases and unit conversions. At that point, everything just clicked and I went back to read through the original documentation and found exactly what we needed."}

\vspace{12pt}

\textit{"My biggest challenge was figuring out how to use the composer files. It was my first time dealing with packages, and Utopia-PHP had a lot of packages to go through. [Ultimately,] I was able to figure out how I should use the composer files for my implementation."}

\subsubsection{Collaborative development}

Collaborative development was discussed in 14 of 25 posts. Students discussed the benefits and difficulties of working with a community of engineers, both social (like communication) and technical (like merge conflicts).

\vspace{12pt}

\textit{"Our mentor recommended that I try to reach out to one of the main contributors to the project to request more documentation, so I found their email on GitHub and asked for more clarification."}

\vspace{12pt}

\textit{"Ever since I asked however, I have never received any feedback from the project managers."}

\vspace{12pt}

\textit{"However, the problem with my previous solution involved many issues which were merge conflicts, failure on a few test cases, and messy commits."}

\subsubsection{Benefits of contribution to users and society}

References to user or societal impact appeared in 10 of 25 posts, though students rarely connected the impact specifically to their own contribution.

\vspace{12pt}

\textit{"This could be helpful to organizations to know how much energy is being used over the course of a month."}

\vspace{12pt}

\textit{"This is important because it provides developers with a framework to use for modern PHP development without having to deal with the hassle and overhead of the larger, more complicated framework."}

\vspace{12pt}

\textit{"Static code analysis is a very helpful tool that will allow developers to spot bugs and other problems within their code without running it."}

\subsection{(RQ2) How thoroughly do the structured blog posts allow students to reflect on their experience?}

\begin{figure}[h]
    \centering
    \caption{Split violin plot of normalized scores for knowledge-gain from reflection survey; higher indicates more knowledge-gain from reflection (n=25)}
    \label{fig:reflection}
    \includegraphics[width=\columnwidth,alt={Split violin plot comparing survey scores between two groups: students who submitted a pull request (PR) and students who did not (No PR), across four outcome categories: Engineering Self, Course Understanding, Growth Areas, and Social Impact.}]{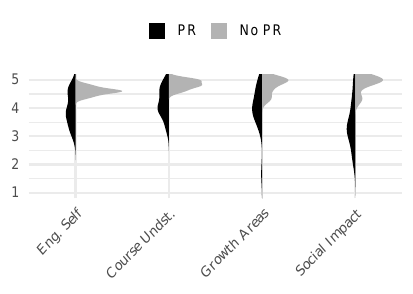}
\end{figure}
\begin{table}[H]
\caption{Normalized scores for knowledge-gain from reflection survey; higher indicates more knowledge-gain from reflection (n=25)}

\label{tab:reflection}
\begin{tabular}{lcccc}
\toprule
\textbf{} & \textbf{\begin{tabular}[c]{@{}c@{}}\small{Cronbach's}\\ $\alpha$\end{tabular}} & \textbf{Mean} & \textbf{SD} \\
\midrule
Engineering Self & 0.79 & 4.10 & 0.57 \\
Course Understanding & 0.75 & 4.34 & 0.53 \\
Areas for Growth & 0.75 & 4.17 & 0.75 \\
Social Impact & 0.71 & 3.72 & 0.95 \\
\bottomrule
\end{tabular}
\end{table}

We recruited a total of 25 students to complete the reflection survey. 20 had successfully completed an open source contribution, and 5 had not. All students published a blog post.

Results are presented in Figure \ref{fig:reflection} and Table \ref{tab:reflection}, normalized to a 1-5 scale. Internal consistency for each construct was assessed using Cronbach's alpha and was found to be acceptable ($> 0.70$). Given the small sample size (n=25), we did not conduct inferential statistical tests.

The results suggest that the structured blog posts were effective at allowing students to gain knowledge into their "engineering self", which Mejia and Turns define as "a student's ability to see themselves as engineers, whether it is in their majors or future careers." This aligned with our primary goal of helping students see how the knowledge they gained from their work on an open source contribution prepared them for their future career. 

To a similar extent, blog posts also seemed to improve students understanding of the course material (which related to collaboration and problem solving strategies) and to identify areas of growth.

The blog posts somewhat helped students see how engineering could have a social impact. While this was not a primary goal, we were surprised to see this was the lowest impact, since most projects had a large impact and the early portion of the blog post asked students to detail that impact.

Though the subgroup sizes are too small to draw reliable conclusions, the data suggest that blog posts may be more useful to students who did not complete a pull request.
\section{Future Work}

Based on conversations with students, one of their biggest challenges was the day-to-day reality of software engineering: finding the right file to edit, fixing formatting errors, waiting for code reviews, and how little time is actually spent writing code. This topic rarely appeared in blog posts, possibly because students viewed these struggles as failures rather than evidence of growth. We may refine the blog post template to elicit more reflection in this area.

This paper focused on reflection's benefits to professional identity development, but we hope to eventually evaluate whether posting on LinkedIn had any impact on students' job prospects.
\section{Limitations}

This study has several limitations that should be considered when interpreting the findings. The sample size is modest and drawn from a single program context, which limits generalizability to other populations and program structures. The program also includes several components beyond the blog post; because we did not include a comparison group or pre-/post-measures, we cannot isolate the effects of blog post writing from the broader program structure. Observed changes in reflection and self-perception may result from the combination of these supports rather than the blog posts alone.
\section{Conclusions}

This study demonstrates that structured blog posts can effectively help students reflect on experiential learning and connect their academic work to professional contexts. Through thematic analysis of student blog posts and quantitative survey data, the blog post framework was particularly successful at helping students develop their "engineering self," their ability to see themselves as engineers prepared for future careers.

Students used the blog posts to identify and articulate problem-solving techniques, celebrate breakthrough moments that built a growth mindset, reflect on collaborative development, and think through the impact of their contribution on users --- four traits that are highly valued by hiring managers. \cite{sahaOpenSourceResume2025} Students' recognition of their skills may be an asset in their efforts to secure employment. 

Our findings also reveal gaps in reflection. Students struggled most with the day-to-day realities of software engineering: navigating codebases, addressing formatting requirements, managing code review, and the limited time spent actually writing code. These challenges rarely surfaced in their blog posts, suggesting the current template does not adequately prompt reflection on these aspects of professional software development.

The intervention achieved a secondary goal of helping students contextualize their open source contributions for potential employers, though we could not evaluate the specific impact due to space constraints. Publishing on LinkedIn allowed students to re-engage their professional networks and present their work in a way that highlighted growth and learning, not just technical output.

Integrating reflective writing into experiential learning programs is a promising way to help students bridge the gap between academic study and professional practice. Structured reflection helps students recognize and articulate the full value of their learning, better preparing them for careers in their chosen fields.

\begin{acks}
This material is based upon work supported by the National Science Foundation under Grant No. 2347311.

\subsubsection*{CRediT author statement}
\textbf{Utsab Saha:} Conceptualization, Formal Analysis, Investigation, Writing – original draft; \textbf{Lola Egherman:} Conceptualization, Writing – review \& editing, Project administration; \textbf{Ramiz Rahman:} Conceptualization, Writing – review \& editing; \textbf{Mohd Toukir Khan:} Conceptualization, Writing – review \& editing; \textbf{Kevin Wang:} Conceptualization, Investigation, Writing – original draft; \textbf{Tyler Menezes:} Conceptualization, Formal Analysis, Investigation, Methodology, Supervision, Writing – original draft
\end{acks}

\bibliographystyle{format/ACM-Reference-Format}
\balance
\bibliography{arxiv}


\begin{thebibliography}{42}


\ifx \showCODEN    \undefined \def \showCODEN     #1{\unskip}     \fi
\ifx \showISBNx    \undefined \def \showISBNx     #1{\unskip}     \fi
\ifx \showISBNxiii \undefined \def \showISBNxiii  #1{\unskip}     \fi
\ifx \showISSN     \undefined \def \showISSN      #1{\unskip}     \fi
\ifx \showLCCN     \undefined \def \showLCCN      #1{\unskip}     \fi
\ifx \shownote     \undefined \def \shownote      #1{#1}          \fi
\ifx \showarticletitle \undefined \def \showarticletitle #1{#1}   \fi
\ifx \showURL      \undefined \def \showURL       {\relax}        \fi
\providecommand\bibfield[2]{#2}
\providecommand\bibinfo[2]{#2}
\providecommand\natexlab[1]{#1}
\providecommand\showeprint[2][]{arXiv:#2}

\bibitem[Braun and Clarke(2006)]%
        {braunUsingThematicAnalysis2006}
\bibfield{author}{\bibinfo{person}{Virginia Braun} {and} \bibinfo{person}{Victoria Clarke}.} \bibinfo{year}{2006}\natexlab{}.
\newblock \showarticletitle{Using thematic analysis in psychology}.
\newblock \bibinfo{journal}{\emph{Qualitative Research in Psychology}} \bibinfo{volume}{3}, \bibinfo{number}{2} (\bibinfo{date}{Jan.} \bibinfo{year}{2006}), \bibinfo{pages}{77--101}.
\newblock
\showISSN{1478-0887}
\href{https://doi.org/10.1191/1478088706qp063oa}{doi:\nolinkurl{10.1191/1478088706qp063oa}}


\bibitem[Braun and Clarke(2021)]%
        {braunOneSizeFits2021}
\bibfield{author}{\bibinfo{person}{Virginia Braun} {and} \bibinfo{person}{Victoria Clarke}.} \bibinfo{year}{2021}\natexlab{}.
\newblock \showarticletitle{One size fits all? {What} counts as quality practice in (reflexive) thematic analysis?}
\newblock \bibinfo{journal}{\emph{Qualitative Research in Psychology}} \bibinfo{volume}{18}, \bibinfo{number}{3} (\bibinfo{date}{July} \bibinfo{year}{2021}), \bibinfo{pages}{328--352}.
\newblock
\showISSN{1478-0887, 1478-0895}
\href{https://doi.org/10.1080/14780887.2020.1769238}{doi:\nolinkurl{10.1080/14780887.2020.1769238}}


\bibitem[Carmack and Heiss(2018)]%
        {carmackUsingTheoryPlanned2018}
\bibfield{author}{\bibinfo{person}{Heather~J. Carmack} {and} \bibinfo{person}{Sarah~N. Heiss}.} \bibinfo{year}{2018}\natexlab{}.
\newblock \showarticletitle{Using the {Theory} of {Planned} {Behavior} to {Predict} {College} {Students}’ {Intent} to {Use} {LinkedIn} for {Job} {Searches} and {Professional} {Networking}}.
\newblock \bibinfo{journal}{\emph{Communication Studies}} \bibinfo{volume}{69}, \bibinfo{number}{2} (\bibinfo{date}{March} \bibinfo{year}{2018}), \bibinfo{pages}{145--160}.
\newblock
\showISSN{1051-0974}
\href{https://doi.org/10.1080/10510974.2018.1424003}{doi:\nolinkurl{10.1080/10510974.2018.1424003}}


\bibitem[Chalikandy(2014)]%
        {chalikandyReflectionToolProfessional2014}
\bibfield{author}{\bibinfo{person}{Muhammed~Ali Chalikandy}.} \bibinfo{year}{2014}\natexlab{}.
\newblock \showarticletitle{Reflection: {A} {Tool} for {Professional} {Development}}.
\newblock \bibinfo{journal}{\emph{Researchers World - International Refereed Social Sciences Journal}} \bibinfo{volume}{5}, \bibinfo{number}{3} (\bibinfo{year}{2014}), \bibinfo{pages}{117--124}.
\newblock
\urldef\tempurl%
\url{https://www.researchersworld.com/index.php/rworld/article/view/859}
\showURL{%
\tempurl}


\bibitem[Cropley(2015)]%
        {cropleyPromotingCreativityInnovation2015}
\bibfield{author}{\bibinfo{person}{David~H. Cropley}.} \bibinfo{year}{2015}\natexlab{}.
\newblock \showarticletitle{Promoting creativity and innovation in engineering education}.
\newblock \bibinfo{journal}{\emph{Psychology of Aesthetics, Creativity, and the Arts}} \bibinfo{volume}{9}, \bibinfo{number}{2} (\bibinfo{year}{2015}), \bibinfo{pages}{161--171}.
\newblock
\showISSN{1931-390X}
\href{https://doi.org/10.1037/aca0000008}{doi:\nolinkurl{10.1037/aca0000008}}


\bibitem[Daniels et~al\mbox{.}(2023)]%
        {danielsLinkedInBlundersMixed2023}
\bibfield{author}{\bibinfo{person}{Ruby~A. Daniels}, \bibinfo{person}{Sara~D. Pemble}, \bibinfo{person}{Danielle Allen}, \bibinfo{person}{Gretchen Lain}, {and} \bibinfo{person}{Leslie~A. Miller}.} \bibinfo{year}{2023}\natexlab{}.
\newblock \showarticletitle{{LinkedIn} {Blunders}: {A} {Mixed} {Method} {Study} of {College} {Students}’ {Profiles}}.
\newblock \bibinfo{journal}{\emph{Community College Journal of Research and Practice}} \bibinfo{volume}{47}, \bibinfo{number}{2} (\bibinfo{date}{Feb.} \bibinfo{year}{2023}), \bibinfo{pages}{90--105}.
\newblock
\showISSN{1066-8926}
\href{https://doi.org/10.1080/10668926.2021.1944932}{doi:\nolinkurl{10.1080/10668926.2021.1944932}}


\bibitem[Dickson and Barr(2019)]%
        {dicksonBringingReflectionComputer2019}
\bibfield{author}{\bibinfo{person}{Paul~E. Dickson} {and} \bibinfo{person}{John Barr}.} \bibinfo{year}{2019}\natexlab{}.
\newblock \showarticletitle{Bringing {Reflection} into {Computer} {Science} {Education}}. In \bibinfo{booktitle}{\emph{Proceedings of the 50th {ACM} {Technical} {Symposium} on {Computer} {Science} {Education}}}. \bibinfo{publisher}{ACM}, \bibinfo{address}{Minneapolis MN USA}, \bibinfo{pages}{1249--1249}.
\newblock
\showISBNx{978-1-4503-5890-3}
\href{https://doi.org/10.1145/3287324.3293733}{doi:\nolinkurl{10.1145/3287324.3293733}}


\bibitem[DuRose and Stebleton(2016)]%
        {duroseLostTranslationPreparing2016}
\bibfield{author}{\bibinfo{person}{Lisa DuRose} {and} \bibinfo{person}{Michael~J. Stebleton}.} \bibinfo{year}{2016}\natexlab{}.
\newblock \showarticletitle{Lost in {Translation}: {Preparing} {Students} to {Articulate} the {Meaning} of a {College} {Degree}}.
\newblock \bibinfo{journal}{\emph{Journal of College and Character}} \bibinfo{volume}{17}, \bibinfo{number}{4} (\bibinfo{date}{Oct.} \bibinfo{year}{2016}), \bibinfo{pages}{271--277}.
\newblock
\showISSN{2194-587X}
\href{https://doi.org/10.1080/2194587X.2016.1230759}{doi:\nolinkurl{10.1080/2194587X.2016.1230759}}


\bibitem[Eggleston et~al\mbox{.}(2022)]%
        {egglestonEmployerStudentMismatch2022}
\bibfield{author}{\bibinfo{person}{Alyson Eggleston}, \bibinfo{person}{Robert Rabb}, {and} \bibinfo{person}{Ronald Welch}.} \bibinfo{year}{2022}\natexlab{}.
\newblock \showarticletitle{Employer and {Student} {Mismatch} in {Early}-{Career} {Skill} {Development}}. In \bibinfo{booktitle}{\emph{2022 {ASEE} {Annual} {Conference} \& {Exposition} {Proceedings}}}. \bibinfo{publisher}{ASEE Conferences}, \bibinfo{address}{Minneapolis, MN}, \bibinfo{pages}{40895}.
\newblock
\href{https://doi.org/10.18260/1-2--40895}{doi:\nolinkurl{10.18260/1-2--40895}}


\bibitem[Feng et~al\mbox{.}(2025)]%
        {fengMultifacetedNatureMentoring2025}
\bibfield{author}{\bibinfo{person}{Zixuan Feng}, \bibinfo{person}{Igor Steinmacher}, \bibinfo{person}{Marco Gerosa}, \bibinfo{person}{Tyler Menezes}, \bibinfo{person}{Alexander Serebrenik}, \bibinfo{person}{Reed Milewicz}, {and} \bibinfo{person}{Anita Sarma}.} \bibinfo{year}{2025}\natexlab{}.
\newblock \showarticletitle{The {Multifaceted} {Nature} of {Mentoring} in {OSS}: {Strategies}, {Qualities}, and {Ideal} {Outcomes}}. In \bibinfo{booktitle}{\emph{2025 {IEEE}/{ACM} 18th {International} {Conference} on {Cooperative} and {Human} {Aspects} of {Software} {Engineering} ({CHASE})}}. \bibinfo{publisher}{IEEE}, \bibinfo{address}{Ottawa, ON, Canada}, \bibinfo{pages}{203--214}.
\newblock
\showISBNx{979-8-3315-3871-2}
\href{https://doi.org/10.1109/CHASE66643.2025.00031}{doi:\nolinkurl{10.1109/CHASE66643.2025.00031}}


\bibitem[French et~al\mbox{.}(2015)]%
        {frenchMountaintopCorporateLadder2015}
\bibfield{author}{\bibinfo{person}{Erica French}, \bibinfo{person}{Janis Bailey}, \bibinfo{person}{Elizabeth van Acker}, {and} \bibinfo{person}{Leigh Wood}.} \bibinfo{year}{2015}\natexlab{}.
\newblock \showarticletitle{From mountaintop to corporate ladder – what new professionals really really want in a capstone experience!}
\newblock \bibinfo{journal}{\emph{Teaching in Higher Education}} \bibinfo{volume}{20}, \bibinfo{number}{8} (\bibinfo{date}{Nov.} \bibinfo{year}{2015}), \bibinfo{pages}{767--782}.
\newblock
\showISSN{1356-2517}
\href{https://doi.org/10.1080/13562517.2015.1085852}{doi:\nolinkurl{10.1080/13562517.2015.1085852}}


\bibitem[George(2002)]%
        {georgeLearningReflectiveJournal2002}
\bibfield{author}{\bibinfo{person}{Susan~E. George}.} \bibinfo{year}{2002}\natexlab{}.
\newblock \showarticletitle{Learning and the reflective journal in computer science}. In \bibinfo{booktitle}{\emph{Proceedings of the twenty-fifth {Australasian} conference on {Computer} science - {Volume} 4}} \emph{(\bibinfo{series}{{ACSC} '02}, Vol.~\bibinfo{volume}{4})}. \bibinfo{publisher}{Australian Computer Society, Inc.}, \bibinfo{address}{AUS}, \bibinfo{pages}{77--86}.
\newblock
\showISBNx{978-0-909925-82-6}


\bibitem[Gorka et~al\mbox{.}(2007)]%
        {gorkaDevelopingRealisticCapstone2007}
\bibfield{author}{\bibinfo{person}{Sandra Gorka}, \bibinfo{person}{Jacob~R. Miller}, {and} \bibinfo{person}{Brandon~J. Howe}.} \bibinfo{year}{2007}\natexlab{}.
\newblock \showarticletitle{Developing realistic capstone projects in conjunction with industry}. In \bibinfo{booktitle}{\emph{Proceedings of the 8th {ACM} {SIGITE} conference on {Information} technology education}} \emph{(\bibinfo{series}{{SIGITE} '07})}. \bibinfo{publisher}{Association for Computing Machinery}, \bibinfo{address}{New York, NY, USA}, \bibinfo{pages}{27--32}.
\newblock
\showISBNx{978-1-59593-920-3}
\href{https://doi.org/10.1145/1324302.1324309}{doi:\nolinkurl{10.1145/1324302.1324309}}


\bibitem[Hang et~al\mbox{.}(2024)]%
        {hangIndustryMentoringInternship2024}
\bibfield{author}{\bibinfo{person}{Kendrick Hang}, \bibinfo{person}{Tyler Schrock}, \bibinfo{person}{Tina~J. Ostrander}, \bibinfo{person}{Roseann Berg}, \bibinfo{person}{Tyler Menezes}, {and} \bibinfo{person}{Kevin Wang}.} \bibinfo{year}{2024}\natexlab{}.
\newblock \showarticletitle{Industry {Mentoring} and {Internship} {Experiences} at a {Community} {College} {Baccalaureate} {Program} in {Software} {Development}}. In \bibinfo{booktitle}{\emph{Proceedings of the 55th {ACM} {Technical} {Symposium} on {Computer} {Science} {Education} {V}. 1}} \emph{(\bibinfo{series}{{SIGCSE} 2024})}. \bibinfo{publisher}{Association for Computing Machinery}, \bibinfo{address}{New York, NY, USA}, \bibinfo{pages}{456--462}.
\newblock
\showISBNx{979-8-4007-0423-9}
\href{https://doi.org/10.1145/3626252.3630878}{doi:\nolinkurl{10.1145/3626252.3630878}}


\bibitem[Hanson et~al\mbox{.}(2024)]%
        {stradaTalentDisrupted2024}
\bibfield{author}{\bibinfo{person}{Andrew Hanson}, \bibinfo{person}{Carlo Salerno}, \bibinfo{person}{Matt Sigelman}, \bibinfo{person}{Mels de Zeeuw}, {and} \bibinfo{person}{Stephen Moret}.} \bibinfo{year}{2024}\natexlab{}.
\newblock \bibinfo{booktitle}{\emph{Talent {Disrupted}}}.
\newblock \bibinfo{type}{{T}echnical {R}eport}. \bibinfo{institution}{Burning Glass Institute and Strada Institute for the Future of Work}. \bibinfo{pages}{1--56} pages.
\newblock
\urldef\tempurl%
\url{https://www.strada.org/reports/talent-disrupted}
\showURL{%
\tempurl}


\bibitem[Initiative(2007)]%
        {OpenSourceDefinition}
\bibfield{author}{\bibinfo{person}{Open~Source Initiative}.} \bibinfo{year}{2007}\natexlab{}.
\newblock \bibinfo{title}{The {Open} {Source} {Definition}}.
\newblock
\urldef\tempurl%
\url{https://opensource.org/osd}
\showURL{%
\tempurl}


\bibitem[Ions and Sutcliffe(2019)]%
        {ionsBarriersConstructingExperiential2019}
\bibfield{author}{\bibinfo{person}{Kevin~John Ions} {and} \bibinfo{person}{Norma Sutcliffe}.} \bibinfo{year}{2019}\natexlab{}.
\newblock \showarticletitle{Barriers to constructing experiential learning claims through reflective narratives: {Student}’s experiences}.
\newblock \bibinfo{journal}{\emph{Higher Education, Skills and Work-Based Learning}} \bibinfo{volume}{10}, \bibinfo{number}{1} (\bibinfo{date}{Sept.} \bibinfo{year}{2019}), \bibinfo{pages}{126--140}.
\newblock
\showISSN{2042-3896}
\href{https://doi.org/10.1108/HESWBL-04-2019-0053}{doi:\nolinkurl{10.1108/HESWBL-04-2019-0053}}


\bibitem[Jackson and Edgar(2019)]%
        {jacksonEncouragingStudentsDraw2019}
\bibfield{author}{\bibinfo{person}{Denise~A Jackson} {and} \bibinfo{person}{Susan Edgar}.} \bibinfo{year}{2019}\natexlab{}.
\newblock \showarticletitle{Encouraging students to draw on work experiences when articulating achievements and capabilities to enhance employability}.
\newblock \bibinfo{journal}{\emph{Australian Journal of Career Development}} \bibinfo{volume}{28}, \bibinfo{number}{1} (\bibinfo{date}{April} \bibinfo{year}{2019}), \bibinfo{pages}{39--50}.
\newblock
\showISSN{1038-4162}
\href{https://doi.org/10.1177/1038416218790571}{doi:\nolinkurl{10.1177/1038416218790571}}


\bibitem[Jackson et~al\mbox{.}(2022)]%
        {jacksonCollaborationToolsDevelopers2022}
\bibfield{author}{\bibinfo{person}{Victoria Jackson}, \bibinfo{person}{André van~der Hoek}, \bibinfo{person}{Rafael Prikladnicki}, {and} \bibinfo{person}{Christof Ebert}.} \bibinfo{year}{2022}\natexlab{}.
\newblock \showarticletitle{Collaboration {Tools} for {Developers}}.
\newblock \bibinfo{journal}{\emph{IEEE Software}} \bibinfo{volume}{39}, \bibinfo{number}{2} (\bibinfo{date}{March} \bibinfo{year}{2022}), \bibinfo{pages}{7--15}.
\newblock
\showISSN{1937-4194}
\href{https://doi.org/10.1109/MS.2021.3132137}{doi:\nolinkurl{10.1109/MS.2021.3132137}}


\bibitem[Kolar and Dickson(2002)]%
        {kolarPreserviceGeneralEducators2002}
\bibfield{author}{\bibinfo{person}{Christine Kolar} {and} \bibinfo{person}{Shirley~V. Dickson}.} \bibinfo{year}{2002}\natexlab{}.
\newblock \showarticletitle{Preservice {General} {Educators}' {Perceptions} of {Structured} {Reflective} {Logs} as {Viable} {Learning} {Tools} in a {University} {Course} on {Inclusionary} {Practices}}.
\newblock \bibinfo{journal}{\emph{Teacher Education and Special Education}} \bibinfo{volume}{25}, \bibinfo{number}{4} (\bibinfo{date}{Oct.} \bibinfo{year}{2002}), \bibinfo{pages}{395--406}.
\newblock
\showISSN{0888-4064}
\href{https://doi.org/10.1177/088840640202500408}{doi:\nolinkurl{10.1177/088840640202500408}}


\bibitem[Kreber(2004)]%
        {kreberAnalysisTwoModels2004}
\bibfield{author}{\bibinfo{person}{Carolin Kreber}.} \bibinfo{year}{2004}\natexlab{}.
\newblock \showarticletitle{An analysis of two models of reflection and their implications for educational development}.
\newblock \bibinfo{journal}{\emph{International Journal for Academic Development}} \bibinfo{volume}{9}, \bibinfo{number}{1} (\bibinfo{date}{May} \bibinfo{year}{2004}), \bibinfo{pages}{29--49}.
\newblock
\showISSN{1360-144X, 1470-1324}
\href{https://doi.org/10.1080/1360144042000296044}{doi:\nolinkurl{10.1080/1360144042000296044}}


\bibitem[Landis and Koch(1977)]%
        {landisMeasurementObserverAgreement1977}
\bibfield{author}{\bibinfo{person}{J.~Richard Landis} {and} \bibinfo{person}{Gary~G. Koch}.} \bibinfo{year}{1977}\natexlab{}.
\newblock \showarticletitle{The {Measurement} of {Observer} {Agreement} for {Categorical} {Data}}.
\newblock \bibinfo{journal}{\emph{Biometrics}} \bibinfo{volume}{33}, \bibinfo{number}{1} (\bibinfo{year}{1977}), \bibinfo{pages}{159--174}.
\newblock
\showISSN{0006-341X}
\href{https://doi.org/10.2307/2529310}{doi:\nolinkurl{10.2307/2529310}}


\bibitem[Langer(2002)]%
        {langerReflectingPracticeUsing2002}
\bibfield{author}{\bibinfo{person}{Arthur~M. Langer}.} \bibinfo{year}{2002}\natexlab{}.
\newblock \showarticletitle{Reflecting on {Practice}: using learning journals in higher and continuing education}.
\newblock \bibinfo{journal}{\emph{Teaching in Higher Education}} \bibinfo{volume}{7}, \bibinfo{number}{3} (\bibinfo{date}{July} \bibinfo{year}{2002}), \bibinfo{pages}{337--351}.
\newblock
\showISSN{1356-2517, 1470-1294}
\href{https://doi.org/10.1080/13562510220144824}{doi:\nolinkurl{10.1080/13562510220144824}}


\bibitem[Lockyer et~al\mbox{.}(2004)]%
        {lockyerKnowledgeTranslationRole2004}
\bibfield{author}{\bibinfo{person}{Jocelyn Lockyer}, \bibinfo{person}{Tunde~S. Gondocz}, {and} \bibinfo{person}{Robert~L. Thivierge}.} \bibinfo{year}{2004}\natexlab{}.
\newblock \showarticletitle{Knowledge translation: {The} role and place of practice reflection}.
\newblock \bibinfo{journal}{\emph{Journal of Continuing Education in the Health Professions}} \bibinfo{volume}{24}, \bibinfo{number}{1} (\bibinfo{year}{2004}), \bibinfo{pages}{50}.
\newblock
\showISSN{0894-1912}
\href{https://doi.org/10.1002/chp.1340240108}{doi:\nolinkurl{10.1002/chp.1340240108}}


\bibitem[Marienau and Fiddler(2002)]%
        {marienauBringingStudentsExperience2002}
\bibfield{author}{\bibinfo{person}{Catherine Marienau} {and} \bibinfo{person}{Morry Fiddler}.} \bibinfo{year}{2002}\natexlab{}.
\newblock \showarticletitle{Bringing {Students}’: {Experience} to the {Learning} {Process}}.
\newblock \bibinfo{journal}{\emph{About Campus}} \bibinfo{volume}{7}, \bibinfo{number}{5} (\bibinfo{date}{Nov.} \bibinfo{year}{2002}), \bibinfo{pages}{13--19}.
\newblock
\showISSN{1086-4822}
\href{https://doi.org/10.1177/108648220200700504}{doi:\nolinkurl{10.1177/108648220200700504}}


\bibitem[Mejia and Turns(2021)]%
        {mejiaCreatingCapacityExplore2021}
\bibfield{author}{\bibinfo{person}{Kenya Mejia} {and} \bibinfo{person}{Jennifer Turns}.} \bibinfo{year}{2021}\natexlab{}.
\newblock \showarticletitle{Creating {Capacity} to {Explore} what {Students} {Learn} from {Reflection} {Activities}: {Validating} the {Knowledge}-gain {Survey}}. In \bibinfo{booktitle}{\emph{2021 {ASEE} {Virtual} {Annual} {Conference} {Content} {Access} {Proceedings}}}. \bibinfo{publisher}{ASEE Conferences}, \bibinfo{address}{Virtual Conference}, \bibinfo{pages}{36872}.
\newblock
\href{https://doi.org/10.18260/1-2--36872}{doi:\nolinkurl{10.18260/1-2--36872}}


\bibitem[Menezes et~al\mbox{.}(2024)]%
        {menezesAIGradingStandupUpdates2024}
\bibfield{author}{\bibinfo{person}{Tyler Menezes}, \bibinfo{person}{Lola Egherman}, {and} \bibinfo{person}{Nikhil Garg}.} \bibinfo{year}{2024}\natexlab{}.
\newblock \showarticletitle{{AI}-{Grading} {Standup} {Updates} to {Improve} {Project}-{Based} {Learning} {Outcomes}}. In \bibinfo{booktitle}{\emph{Proceedings of the 2024 on {Innovation} and {Technology} in {Computer} {Science} {Education} {V}. 1}} \emph{(\bibinfo{series}{{ITiCSE} 2024})}. \bibinfo{publisher}{Association for Computing Machinery}, \bibinfo{address}{New York, NY, USA}, \bibinfo{pages}{17--23}.
\newblock
\showISBNx{979-8-4007-0600-4}
\href{https://doi.org/10.1145/3649217.3653541}{doi:\nolinkurl{10.1145/3649217.3653541}}


\bibitem[Menezes et~al\mbox{.}(2022)]%
        {menezesOpenSourceInternshipsIndustry2022}
\bibfield{author}{\bibinfo{person}{Tyler Menezes}, \bibinfo{person}{Alexander Parra}, {and} \bibinfo{person}{Mingjie Jiang}.} \bibinfo{year}{2022}\natexlab{}.
\newblock \showarticletitle{Open-{Source} {Internships} {With} {Industry} {Mentors}}. In \bibinfo{booktitle}{\emph{Proceedings of the 27th {ACM} {Conference} on on {Innovation} and {Technology} in {Computer} {Science} {Education} {Vol}. 1}} \emph{(\bibinfo{series}{{ITiCSE} '22})}. \bibinfo{publisher}{Association for Computing Machinery}, \bibinfo{address}{New York, NY, USA}, \bibinfo{pages}{365--371}.
\newblock
\showISBNx{978-1-4503-9201-3}
\href{https://doi.org/10.1145/3502718.3524763}{doi:\nolinkurl{10.1145/3502718.3524763}}


\bibitem[Narayanan et~al\mbox{.}(2023)]%
        {narayananScalableApproachSupport2023}
\bibfield{author}{\bibinfo{person}{Sathya Narayanan}, \bibinfo{person}{Leslie Maxwell}, \bibinfo{person}{Mariana~Anita Garcia}, \bibinfo{person}{Utsab Saha}, {and} \bibinfo{person}{Tyler Menezes}.} \bibinfo{year}{2023}\natexlab{}.
\newblock \showarticletitle{A scalable approach to support computer science students in their learning and preparation as software engineers}. In \bibinfo{booktitle}{\emph{{IEEE} {Frontiers} in {Education} 2023}}. \bibinfo{publisher}{IEEE Press}, \bibinfo{address}{College Station, TX, USA}, \bibinfo{pages}{1--5}.
\newblock
\href{https://doi.org/10.1109/FIE58773.2023.10343322}{doi:\nolinkurl{10.1109/FIE58773.2023.10343322}}


\bibitem[Resch et~al\mbox{.}(2022)]%
        {reschAnalysisEffectAnswering2022}
\bibfield{author}{\bibinfo{person}{Cheryl Resch}, \bibinfo{person}{Patriel Stapleton}, \bibinfo{person}{Benjamin Rheault}, \bibinfo{person}{Amy Wu}, {and} \bibinfo{person}{Christina GaRdner-Mccune}.} \bibinfo{year}{2022}\natexlab{}.
\newblock \showarticletitle{Analysis of {Effect} of {Answering} {Reflection} {Prompts} in a {Computer} {Organization} {Class}}. In \bibinfo{booktitle}{\emph{2022 {ASEE} {Annual} {Conference} \& {Exposition} {Proceedings}}}. \bibinfo{publisher}{ASEE Conferences}, \bibinfo{address}{Minneapolis, MN}, \bibinfo{pages}{40428}.
\newblock
\href{https://doi.org/10.18260/1-2--40428}{doi:\nolinkurl{10.18260/1-2--40428}}


\bibitem[Robertson(2005)]%
        {robertsonReflectionProfessionalPractice2005}
\bibfield{author}{\bibinfo{person}{Kathryn Robertson}.} \bibinfo{year}{2005}\natexlab{}.
\newblock \showarticletitle{Reflection in professional practice and education}.
\newblock \bibinfo{journal}{\emph{Australian Family Physician}} \bibinfo{volume}{34}, \bibinfo{number}{9} (\bibinfo{date}{Sept.} \bibinfo{year}{2005}), \bibinfo{pages}{781--783}.
\newblock
\showISSN{0300-8495}


\bibitem[Rogers(2001)]%
        {rogersReflectionHigherEducation2001}
\bibfield{author}{\bibinfo{person}{Russell~R. Rogers}.} \bibinfo{year}{2001}\natexlab{}.
\newblock \showarticletitle{Reflection in {Higher} {Education}: {A} {Concept} {Analysis}}.
\newblock \bibinfo{journal}{\emph{Innovative Higher Education}} \bibinfo{volume}{26}, \bibinfo{number}{1} (\bibinfo{date}{Sept.} \bibinfo{year}{2001}), \bibinfo{pages}{37--57}.
\newblock
\showISSN{0742-5627, 1573-1758}
\href{https://doi.org/10.1023/A:1010986404527}{doi:\nolinkurl{10.1023/A:1010986404527}}


\bibitem[Saha et~al\mbox{.}(2025)]%
        {sahaOpenSourceResume2025}
\bibfield{author}{\bibinfo{person}{Utsab Saha}, \bibinfo{person}{Jeffrey D'Andria}, {and} \bibinfo{person}{Tyler Menezes}.} \bibinfo{year}{2025}\natexlab{}.
\newblock \bibinfo{title}{The {Open} {Source} {Resume}: {How} {Open} {Source} {Contributions} {Help} {Students} {Demonstrate} {Alignment} with {Employer} {Needs}}.
\newblock
\href{https://doi.org/10.48550/ARXIV.2510.25180}{doi:\nolinkurl{10.48550/ARXIV.2510.25180}}
\newblock
\shownote{Version Number: 1}.


\bibitem[Samuels and Betts(2007)]%
        {samuelsCrossingThresholdDescription2007}
\bibfield{author}{\bibinfo{person}{Mary Samuels} {and} \bibinfo{person}{Jan Betts}.} \bibinfo{year}{2007}\natexlab{}.
\newblock \showarticletitle{Crossing the threshold from description to deconstruction and reconstruction: using self‐assessment to deepen reflection}.
\newblock \bibinfo{journal}{\emph{Reflective Practice}} \bibinfo{volume}{8}, \bibinfo{number}{2} (\bibinfo{date}{May} \bibinfo{year}{2007}), \bibinfo{pages}{269--283}.
\newblock
\showISSN{1462-3943}
\href{https://doi.org/10.1080/14623940701289410}{doi:\nolinkurl{10.1080/14623940701289410}}


\bibitem[Sarig(2005)]%
        {sarigFosteringReflectiveWriting2005}
\bibfield{author}{\bibinfo{person}{Gissi Sarig}.} \bibinfo{year}{2005}\natexlab{}.
\newblock \showarticletitle{Fostering {Reflective} {Writing} by {Structuring} {Writing}-{To}-{Learn} {Tasks}}.
\newblock In \bibinfo{booktitle}{\emph{Effective {Learning} and {Teaching} of {Writing}: {A} {Handbook} of {Writing} in {Education}}}, \bibfield{editor}{\bibinfo{person}{Gert Rijlaarsdam}, \bibinfo{person}{Huub van~den Bergh}, {and} \bibinfo{person}{Michel Couzijn}} (Eds.). \bibinfo{publisher}{Springer Netherlands}, \bibinfo{address}{Dordrecht}, \bibinfo{pages}{499--517}.
\newblock
\showISBNx{978-1-4020-2739-0}
\urldef\tempurl%
\url{https://doi.org/10.1007/978-1-4020-2739-0_34}
\showURL{%
\tempurl}


\bibitem[Sudirman et~al\mbox{.}(2024)]%
        {sudirmanReinforcingReflectivePractice2024}
\bibfield{author}{\bibinfo{person}{Anselmus Sudirman}, \bibinfo{person}{Adria~Vitalya Gemilang}, \bibinfo{person}{Thadius Marhendra~Adi Kristanto}, \bibinfo{person}{Rr.~Hasti Robiasih}, \bibinfo{person}{Isti’anatul Hikmah}, \bibinfo{person}{Andhi~Dwi Nugroho}, \bibinfo{person}{J.~C.~Setyo Karjono}, \bibinfo{person}{Titi Lestari}, \bibinfo{person}{Theresia~Laksmi Widyarini}, \bibinfo{person}{Afria~Dian Prastanti}, \bibinfo{person}{Moh.~Rusnoto Susanto}, {and} \bibinfo{person}{Burhanudin Rais}.} \bibinfo{year}{2024}\natexlab{}.
\newblock \showarticletitle{Reinforcing {Reflective} {Practice} through {Reflective} {Writing} in {Higher} {Education}: {A} {Systematic} {Review}}.
\newblock \bibinfo{journal}{\emph{International Journal of Learning, Teaching and Educational Research}} \bibinfo{volume}{23}, \bibinfo{number}{5} (\bibinfo{date}{May} \bibinfo{year}{2024}), \bibinfo{pages}{454--474}.
\newblock
\showISSN{16942493, 16942116}
\href{https://doi.org/10.26803/ijlter.23.5.24}{doi:\nolinkurl{10.26803/ijlter.23.5.24}}


\bibitem[Swain(1998)]%
        {swainStudyingTeachersTransformations1998}
\bibfield{author}{\bibinfo{person}{Sherry~Seale Swain}.} \bibinfo{year}{1998}\natexlab{}.
\newblock \showarticletitle{Studying {Teachers}' {Transformations}: {Reflection} as {Methodology}}.
\newblock \bibinfo{journal}{\emph{The Clearing House: A Journal of Educational Strategies, Issues and Ideas}} \bibinfo{volume}{72}, \bibinfo{number}{1} (\bibinfo{date}{Sept.} \bibinfo{year}{1998}), \bibinfo{pages}{28--34}.
\newblock
\showISSN{0009-8655}
\href{https://doi.org/10.1080/00098659809599381}{doi:\nolinkurl{10.1080/00098659809599381}}


\bibitem[Towers et~al\mbox{.}(2011)]%
        {towersStudentsPerceptionsEngineering2011}
\bibfield{author}{\bibinfo{person}{Emily Towers}, \bibinfo{person}{Jennifer~A. Simonovich}, {and} \bibinfo{person}{Yevgeniya~V. Zastavker}.} \bibinfo{year}{2011}\natexlab{}.
\newblock \showarticletitle{Students' perceptions of the engineering profession and implications for interest in the field}. In \bibinfo{booktitle}{\emph{2011 {Frontiers} in {Education} {Conference} ({FIE})}}. \bibinfo{publisher}{IEEE}, \bibinfo{pages}{F3D--1--F3D--7}.
\newblock
\href{https://doi.org/10.1109/FIE.2011.6142960}{doi:\nolinkurl{10.1109/FIE.2011.6142960}}


\bibitem[Turns et~al\mbox{.}(2014)]%
        {turnsIntegratingReflectionEngineering2014}
\bibfield{author}{\bibinfo{person}{Jennifer~A. Turns}, \bibinfo{person}{Brook Sattler}, \bibinfo{person}{Ken Yasuhara}, \bibinfo{person}{Jim~L. Borgford-Parnell}, {and} \bibinfo{person}{Cynthia~J. Atman}.} \bibinfo{year}{2014}\natexlab{}.
\newblock \showarticletitle{Integrating {Reflection} into {Engineering} {Education}}. \bibinfo{publisher}{ASEE Conferences}, \bibinfo{pages}{24.776.1--24.776.16}.
\newblock
\href{https://doi.org/10.18260/1-2--20668}{doi:\nolinkurl{10.18260/1-2--20668}}


\bibitem[Ugur and Çakiroglu(2024)]%
        {ugurImplementingReflectiveThinking2024}
\bibfield{author}{\bibinfo{person}{Nursel Ugur} {and} \bibinfo{person}{Ünal Çakiroglu}.} \bibinfo{year}{2024}\natexlab{}.
\newblock \showarticletitle{Implementing {Reflective} {Thinking} in {Computer} {Science} {Unplugged} to {Enhance} {Computational} {Thinking}}.
\newblock \bibinfo{journal}{\emph{International Journal of Technology in Education and Science}} \bibinfo{volume}{8}, \bibinfo{number}{2} (\bibinfo{date}{April} \bibinfo{year}{2024}), \bibinfo{pages}{196--218}.
\newblock
\showISSN{2651-5369}
\href{https://doi.org/10.46328/ijtes.515}{doi:\nolinkurl{10.46328/ijtes.515}}


\bibitem[Wiggins and McTighe(2008)]%
        {wigginsUnderstandingDesign2008}
\bibfield{author}{\bibinfo{person}{Grant Wiggins} {and} \bibinfo{person}{Jay McTighe}.} \bibinfo{year}{2008}\natexlab{}.
\newblock \bibinfo{booktitle}{\emph{Understanding {By} {Design}}}.
\newblock \bibinfo{publisher}{Assn. for Supervision \& Curriculum Development}, \bibinfo{address}{Alexandria, Va}.
\newblock
\showISBNx{978-1-4166-0035-0}


\bibitem[Zarestky et~al\mbox{.}(2022)]%
        {zarestkyReflectiveWritingSupports2022}
\bibfield{author}{\bibinfo{person}{Jill Zarestky}, \bibinfo{person}{Michelle Bigler}, \bibinfo{person}{Mollie Brazile}, \bibinfo{person}{Tobin Lopes}, {and} \bibinfo{person}{Wolfgang Bangerth}.} \bibinfo{year}{2022}\natexlab{}.
\newblock \showarticletitle{Reflective {Writing} {Supports} {Metacognition} and {Self}-regulation in {Graduate} {Computational} {Science} and {Engineering}}.
\newblock \bibinfo{journal}{\emph{Computers and Education Open}}  \bibinfo{volume}{3} (\bibinfo{date}{Dec.} \bibinfo{year}{2022}), \bibinfo{pages}{100085}.
\newblock
\showISSN{26665573}
\href{https://doi.org/10.1016/j.caeo.2022.100085}{doi:\nolinkurl{10.1016/j.caeo.2022.100085}}


\end{thebibliography}

\end{document}